# THE DESIGNING OF ONLINE MULTIPLE INTELLIGENCE TOOLS FOR LECTURERS AT POLYTECHNIC

Sazilah Salam, Siti Nurul Mahfuzah Mohamad,
Norasiken Bakar

Faculty of Information and Communication Technology
Universiti Teknikal Malaysia Melaka (UTeM)
Malacca, Malaysia
sazilah@utem.edu.my, fuza1112@yahoo.com,
norasiken@utem.edu.my

Linda Khoo Mei Sui

Centre for Languages and Human Development,
Universiti Teknikal Malaysia Melaka (UTeM)
Malacca, Malaysia
linda@utem.edu.my

*Abstract*— **This paper addresses the designing of Online Multiple Intelligence (MI) Teaching Tools for Polytechnic lecturers. These teaching tools can assist lecturers to create their own teaching materials without having any knowledge of Information Technology (IT) especially in programming. The theory of MI is used in this paper and this theory postulates that everybody has at least two or more intelligences. Multiple approaches embedded into a series of activities via online teaching tools must be implemented in order to achieve effective teaching and learning in the classroom. The objectives of this paper are to identify the relationship between the students self-perceived MI and their academic achievement in Polytechnic, and design online MI tools for teaching at Polytechnic. This paper also addressed the theoretical framework and MI teaching activities. The instrument used for this study was *Ujian Multiple Intelligence* (UMI). The results showed Polytechnic students have strength in Interpersonal, Visual-Spatial and Verbal-Linguistic intelligences.**

**Keywords-** *Multiple Intelligences, Online Teaching Tools, Interpersonal, Visual-Spatial, Verbal-Linguistic*

## I. INTRODUCTION

Multiple Intelligences theory was introduced by Howard Gardner (1983) and generally, there are nine types of intelligence that may be owned by someone. According to Gardner (2006), every human being has at least two or more intelligences such as verbal linguistic, logic mathematic, intrapersonal, interpersonal, visual spatial, musical, natural and existential intelligences. It is believed that every human being has at least one intelligent, and some of them can even possess to a maximum of eight intelligences. This theory can be applied in the teaching and learning process where suitable teaching materials can be prepared by lecturers to cater to the strengths of their students in the classroom. Lecturers should allow considerable elements of students' choice when designing activities and tasks for the intelligences because students would perform well in the tasks which appeal to their interests. They can allow or should encourage students to explore and learn on their own, and they can facilitate students to direct their own learning. Lecturers can also help students to understand and recognize their strengths, and relate to real-life activities that would enhance their learning.

Everyone is born possessing the nine intelligences. However, all students will come into the classroom with different sets of intelligences. This means that each student will have his or her own unique set of intellectual strengths and weaknesses. Without the MI theory in the classroom, students would not have the opportunity to explore their own intelligences. Indirectly, this will have an impact on students' achievement and thus reduce their attention and interest in the classroom. With the aids of teaching tools based on MI, many studies showed that they can increase students' confidence and improve their academic achievement. An awareness of MI theory has urged lecturers to find ways to help their students. Therefore, an online MI teaching tools is designed to facilitate lecturers to create interactive course materials or teaching templates based on MI theory. This study only designs an online MI teaching tool that supports MI activities and tasks for polytechnic lecturers. The teaching activities were limited to three intelligences only. These intelligences are Interpersonal, Visual-Spatial and Verbal-Linguistic intelligences.

### A. Problem Statement

Many research related to MI theory indicate that students who apply MI theory in their learning contributed significant differences in their learning output. Lecturers should plan in a way that can involve as many of the intelligences as possible (Ghazi, Shahzada, Gilani, Shabbir and Rashid, 2011; Moran, Kornhaber and Gardner, 2006; Rettig, 2005) because all the intelligences contributes to the student achievements (Parviz & Farhady, 2012; Hajimirzayee & Abadi, 2012; Mostafari, Akbari & Masoominezhad, 2012). Most of the research conducted on MI in teaching and learning have yielded mixed results. Certain studies shown that teaching students about the strength of using MI in learning gained many benefits while other studies claimed that there is a cause and effect between intelligence and academic achievement (Ganggi, 2011; Laidra, Pullmann & Allik; 2007, Waterhouse, 2006).

Past experiments showed the teaching effectiveness and efficiency are still limited. Thus, Ganggi, (2011) recommended that extended research need to be carried out to investigate the best practices to apply MI theory into teaching practice. According to Tyler (2011), the MI theory can be applied into institutions of higher education and pose challenges for instructors to apply these intelligences into their teaching. Recent research in education also indicates that there is a lack of creative and innovative teaching strategies among the teachers (Che Mah Yusof & Mariani, 2001).

This study fills in the gap of the teaching tools used by Polytechnic lecturers. To define a research problem, interview and survey questionnaire have been piloted to a Polytechnic lecturers in Malaysia. An online learning used in teaching and learning for Polytechnic lecturers is called as Curriculum Information Document Online System (CIDOS). The objectives of the survey are to understand the use of e-learning (CIDOS) in teaching and learning at Polytechnic and identify the critical subjects that have problems of delivery to students. The focus will concentrate on how the lecturers prepare and design their teaching materials. The major weakness found in Polytechnic was the technology tool used in teaching and learning process. The result from the survey showed that most of the lecturers commented that the e-learning is not user friendly, too complex and difficult to use. CIDOS also did not fit into their teaching methodology and incompatible with a few courses.

*B. Project Objective*

The objectives of this paper are:
1. To identify the relationship between the students self-perceived Multiple Intelligences and their academic achievement in Polytechnic.
2. To design Online Multiple Intelligences Tool for Teaching at Polytechnic.

*C. Research Question*

Based on the problem background and research objectives of this project, five research questions were derived:
1. What are the relationships between the students self-perceived Multiple Intelligences and their academic achievement in Polytechnic?
2. What tools are currently available for teaching using online?
3. Which intelligences are appropriate for this study?
4. What are the best teaching tools to be considered for each intelligence?
5. How to design the teaching tools based on the intelligences selected?

*D. Scope of research*

An online MI tool for teaching is designed as a teaching tool to overcome the current problem and establish a teaching method in Polytechnic which hope to improve the interaction among students. Learning contents were based on FP201 Programming Fundamentals that covered the sub topics of function and pointer in C-Programming and the target respondents were Polytechnic lecturers and Diploma students.

## II. LITERATURE REVIEW

Howard Gardner (1993) is a psychologist and professor at Harvard University's Graduate School of Education. Based on his study, Gardner developed the theory of multiple intelligences. Each person has at least two or three dominant intelligences that he or she uses to complete daily tasks, solve problems and respond in stressful situations. In addition, most people have the ability to develop skills in each of the intelligences and learn through them. In his theory, Gardner defines intelligence as the "biopsychological potential to process information that can be activated in a cultural setting to solve problems or create products that are of value in a culture" (Gardner, 2000). This definition implies that depending on the setting or domain, different intelligences are used to solve problems and fashion products such as compositions, music or poetry. It is also important to note that Gardner considers intelligences as something that cannot be seen or measured but looking into potentials that will or will not be activated depending upon the values of a particular culture and the opportunities available in that culture. According to MI theory, types of learning styles are as follows:

*i- Visual-Spatial*
Visual-Spatial is the ability to perceive and recreate the visual world accurately, to visualize in one's head and give some kind of order and meaning to objects in space. For this intelligence, students should be able to demonstrate visual perception, which would include the use of images, designs, colours, pictures, visual symbols, patterns designs and shapes. Drawing and painting pictures have to be considered in this category, such as drawing a map. Instructors can employ the use of visual-spatial learning environment equipped with access to visual tools, intentional display areas and changing perspectives through rotating seating (Campbell, Campbell, & Dickinson, 1996).

*ii- Verbal-Linguistic*
Verbal-Linguistic intelligence is the ability to understand and use language, both written and spoken, sensitivity to the meaning of words and the different functions of language. This intelligence is the most common used in daily communication, whether formal or informal, written or spoken. Some of the activities that facilitate the development of this intelligence include reading, vocabulary, writing and making speeches, journal or diary keeping, creative and poetry writing, debates, impromptu speaking, or story telling (Lazear, 1994).

*iii- Logical-Mathematical*

Logical-Mathematical intelligence is the ability to use inductive and deductive thinking, numbers and abstract patterns. This intelligence is often referred to as scientific thinking such as comparing, contrasting and synthesizing information. We often use logical-mathematical intelligence in our daily lives in activities such as making shopping lists and budgeting. All forms of problem solving come under this category. To include the mathematical-logical intelligence, Campbell, Campbell and Dickinson (1996) offer many great ideas such as diverse questioning strategies, posing open-ended problems, applying math to real world situations and using concrete objects to demonstrate understanding.

*iv- Bodily-Kinesthetic*

Bodily-kinesthetic is the ability to use and understand physical movement, a mastery over body movement or the ability to manipulate objects with finesse. For bodily/ kinesthetic intelligence, the emphasis is on practical demonstration or action such as physical exercises, sports, games, martial arts and drama. Students should be able to demonstrate control of various motor activities through activities like games, athletics and exercise, dance, drama, gestures and mime. To include those students who have many strengths with "hands-on" material, instructors might try using creative movement, hands-on thinking, field trips, classroom theatre, competitive and cooperative games, use of kinesthetic imagery, tactile materials and experiences and using communicative body language (Armstrong, 1994; Campbell, Campbell, & Dickinson, 1996).

*v- Musical*

Musical Intelligence is the ability to discern meaning in or to communicate with tonal patterns, sounds, rhythms and beats. Musical intelligence calls for students to display auditory skills, which basically includes hearing and producing sounds. To incorporate the musical intelligence, instructors can play mood and background music, linking tunes with class concepts and giving students musical options for their projects or assignments (Armstrong, 1994). Campbell, Campbell, & Dickinson (1996) suggested background and mood music aided to set an engaging climate for students to work in, as well as providing supportive technology. Thus, the decision to include a portable compact disc player in a class has given an option to students.

*vi- Naturalist*

Naturalistic is the most recent addition to Gardner's theory (Gardner, 2001) and met with more resistance than his original seven intelligences. According to Gardner, individuals who are high in this type of intelligence are more in tune with nature and often found the interest in nurturing, exploring the environment and learning about other species.

*vii- Interpersonal*

Interpersonal intelligence is the ability to make distinctions among other individuals with regard to their moods, motivations and temperaments as well as to communicate with others. Interpersonal intelligence would emphasize the ability of students to work as part of a group, which requires verbal and non-verbal communication skills, co-operation and empathy in within a group. The focus would be on learning in groups or in pairs. Students should be encouraged to use the knowledge and skills to help their groups and peers to complete tasks given successfully. To help students learn with and from others, instructors can incorporate cooperative groups, interpersonal interaction, conflict mediation, peer teaching, group brainstorming, peer sharing, community involvement, and parties or social gatherings as context for learning (Armstrong, 1994).

*viii-Intrapersonal.*

The person with intrapersonal intelligence can be introverted, prefers to work alone and has clear knowledge of what he or she needs in most settings. Some people who agree with Gardner's theories believe that those who possess intrapersonal intelligence in great degree need opportunities to work alone, but may require some extra care because of a high level of perfectionism associated with this form of intelligence. Instructors can include the intrapersonal intelligence through activities such as independent study, self-paced instruction, individualized projects and games, private spaces for study, one minute reflection period, encourage personal connections, options for assignments or projects, exposure to inspirational/ motivational curricula, journal keeping, self-esteem activities, and goal setting (Armstrong, 1994).

In general, MI also helps teachers create more personalized and diverse lessons to accommodate their students learning needs, which leads to more opportunities for students to learn the expected material (Wilson, 1998). There are various ways to implement the MI theory and the implementation may look different in every classroom (Baum, Vines, & Slatin, 2005). There is a study showed that there was a significant positive correlation between perceived linguistic, logic-mathematic and visual spatial intelligence academic achievement of the study while the relationship between self-perceived bodily/kinesthetic and musical intelligence and academic achievement was very weak (Ghazi et al., 2011).

The findings found in a another research showed that Logic Mathematic group had the highest score on the post test and learning gained followed by visual spatial and verbal linguistic intelligences (Bushro & Halimah, 2007). In related study, the dominant intelligence of students for both experiment and control groups, before and after treatment, showed the element of logical-mathematical intelligence which includes analytical skills as well as logical thinking ability (Ozdemir, Giineysu & Tekkaya, 2006).

## III. METHODOLOGY

This part described the research methodology and the research design of this study. This part also addressed aspects of research design such as theoretical framework and proposed multiple intelligences teaching activities.

### A. Interview

The purpose of the interview was to gather information about the current problem in teaching and learning using e-learning CIDOS at Malaysia Polytechnic. The lecturers were randomly selected in this study. This interview approach provides to obtain in depth information, easy analyzing and more systematic and comparable information from different individuals. In addition to pre-prepared questions, it aims to receive more detailed information on all aspects related to the research problem.

### B. Questionnaire

The survey questionnaire helped to determine the usage and current problem of e-learning – CIDOS at Polytechnic as well as identify the lecturer's needs in preparing the teaching materials and critical subjects. The instrument used a Likert Scale (1-5). The questionnaires and instructions were given to respondents. Participation was voluntary. The questionnaires were personally administered to respondents from *Politeknik Ungku Omar, Politeknik Seberang Perai, Politeknik Merlimau* and *Politeknik Shah Alam.* For the Internet-based survey, the questionnaires were distributed via e-mail to the respondents and social networking.

### C. Research Instrument

There are many ways to determine students' intelligence strengths. Several inventories, questionnaires, and tests had been created for this purpose. The UMI is used to determine which intelligences are the strongest for Polytechnic student. The form was taken from (Bushro, 2008). This form was translated in Malay in order to make it easier and understandable to the students. Students were asked to take the MI Test using online application. Once the items of the survey instrument were scored, the points for each of the intelligences were totaled for each student using the Ms. Excel.

### D. Theoretical Framework

Figure 1 illustrated the research framework of this paper. Exclusively, the research framework defined the teaching methods as independent variable, multiple intelligences as moderating variables and teaching outcome as dependent variable. Traditionally teaching aids are still used among the lecturers at Polytechnic. In general, this traditional method only gives benefit to students with verbal-linguistic intelligence where lecturers prefer to distribute handouts and teaching using the slide presentation. It is mentioned in the study by Aris, Abu, Ellington & Dhamotharan (2000), lecturers are more likely using speech and chalk, charts and posters. As a result, students easily bored in the classroom due to the practice of the traditional method.

In line with the development of IT, some lecturers have started to use teaching aids online. With the exposure and courses at the departmental level, in particular young lecturers who have already familiarized with teaching and learning online. Many studies have shown the benefits of using online teaching and learning which increase productivity and reduce time in the preparation of teaching materials. The online facility enables lecturer to prepares notes, tutorials, quizzes faster and easier. This is in accordance with the opinion of Toth, Audrey & Foulger (2010) who states that the shift from traditional teaching to technology-based instruction produces a positive effect on students.

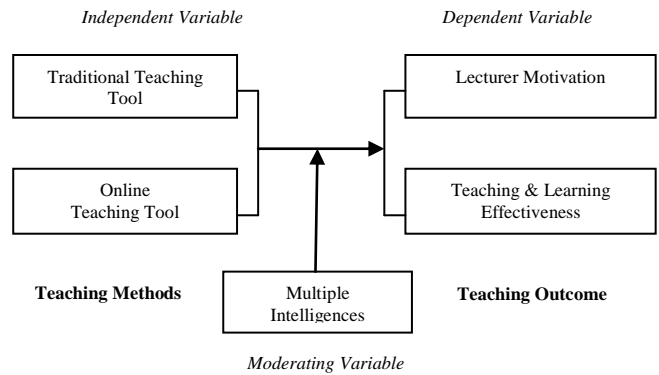

Figure 1. Theoretical Framework

The dependent variable defined in this research framework was the teaching outcome. Teaching outcome will be measured through lecturer motivation survey and teaching and learning questionnaire. Motivation is the drive to the will and desire to succeed or achieve something. According to Woolfolk (1990), motivation refers to an internal power rise, directing and controlling interest and human behavior. The hypothesis for motivation in this treatment is lecturer feel the encouragement to use the online MI teaching tools in classroom. The moderator variable of this research is the theory of MI.

### E. Proposed Online MI Activities

Lecturers become curriculum developers either to create or design teaching materials. MI Teaching Tools are a rich collection of teaching templates designed to save lecturers time in preparing their teaching materials. Most of the free online teaching tools offer the trial version and user cannot edit and save the templates. Using the MI Teaching Tools, lecturers can create free games, quizzes, activities, interactive diagrams and applications based on the MI theory in a few seconds. Not only that, no coding is required because these tools are easy to use and user friendly. The tools also offer the user to organize, save, print and export to other document such as jpg file, doc and etc. In addition, the templates can be changed later or edited at any time. This tool actually gives lecturers the power to generate their own in-depth, effective and innovative teaching tools.

*i. Interpersonal Teaching Activities*

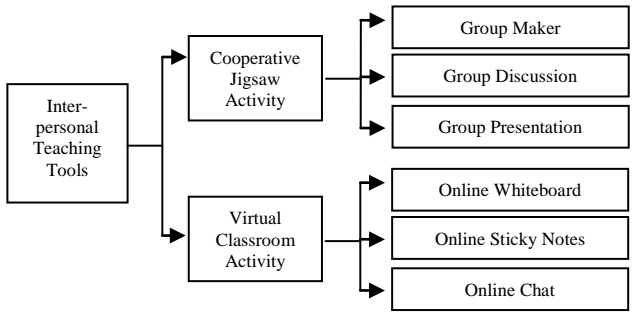

Figure 2. Interpersonal Teaching Activities

Collaborative learning can be defined as a method of teaching in which students work in groups of two or more, mutually searching for understanding, solutions, or meanings or creating a product (McInnerney & Roberts, 2004; Smith & MacGregor, 1992). Collaboration is also referred as shared activity of the students, interaction between students, or participating in learning communities, where the participants are committed to or engaged in shared goals and problem solving (Arvaja, Hakkinen, & Kankaanranta, 2008).

This study implemented jigsaw technique as collaborative learning activity. Therefore, the interface was designed based on students learning style to support jigsaw technique activity. One of the tool offers is the Group Maker. Group Maker tool lets the lecturer generate groups of two, three, four and five students in seconds. This tool can be used easily in class. These groups were in random and can be shuffled over and over as needed. Figure 2 showed the interpersonal teaching activities used in this study.

*ii. Visual Spatial Teaching Activities*

There are few visual spatial teaching activities provided in this study (Figure 4). One of the tool offers is the Interactive Diagram Generator. This tool allow lecturer to create their own interactive diagrams just by adding own content. Interactive Diagram is dynamic teaching tools that allow lecturers to make and customize their notes. The user just adds the word at each box in the diagram where each box is fully editable enabling. Besides that, user can also change the background color, font color and font formatting. Using this generator, lecturers can use to make topic summaries, construct notes for revision, plan teaching activities and develop pre-prepared colour-coded notes and ideas. This tool also provides mind map maker which is lecturers can create their own colourful mind map.

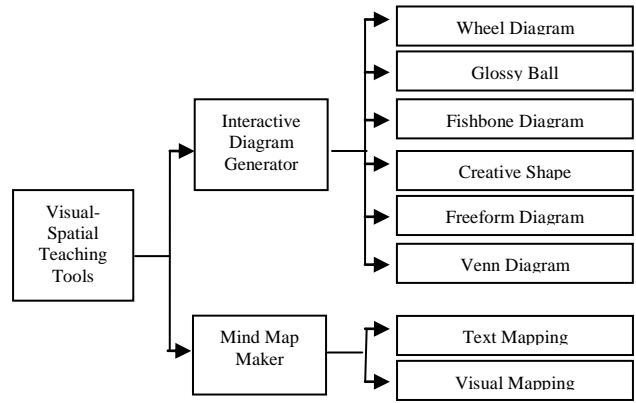

Figure 3. Visual Spatial Teaching Activities

*iii. Verbal-Linguistic Teaching Activities*

There are various verbal-linguistic activities that lecturer can use in the classroom. These intelligence activities will be great for motivating students who are *Word Smart*. Students are *word smart* when they are good at reading, using text and thinking. This tool provides Quiz Maker and Word Game Maker that based on verbal-linguistic intelligence (Figure 4). For example, Multiple Choice Quiz Maker allows anyone to generate their own Flash-based multiple choice quizzes without the need for programming skills. This tool is easy to use because lecturers just need to key in the question and answer into the template, make a selection and the quiz is then instantly generated.

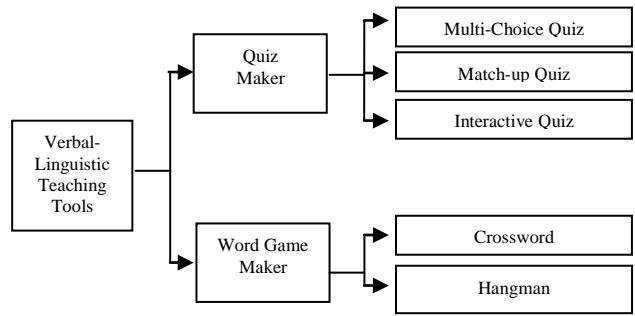

Figure 4. Verbal Linguistic Teaching Activities

## IV. RESULTS

The collected data was entered into SPSS-19 and was analyzed using appropriate statistical tests. The central tendency and variability of the MI of the sampled students was measured using Mean and SD respectively. MI and academic achievement scores were correlated using

Pearson's product moment Correlation. The relationship between MI and CGPA was a positive correlation (Table 1).

Table 1. UMI Result

| No | INTELLIGENCE | r | p | RESULT |
|---|---|---|---|---|
| 1 | Interpersonal | 0.678 | 0.000 | There is a significant positive correlation between perceived interpersonal intelligence and academic achievement of the students. It can also be concluded that this correlation is moderate. |
| 2 | Verbal-Linguistic | 0.471 | 0.027 | There is a significant positive correlation between perceived verbal-linguistic intelligence and academic achievement of the students. It can also be concluded that this correlation is moderate. |
| 3 | Bodily-Kinesthetic | 0.438 | 0.000 | There is a significant positive correlation between perceived Bodily-Kinesthatic intelligence and academic achievement of the students. It can also be concluded that this correlation is moderate. |
| 4 | Visual-Spatial | 0.408 | 0.003 | There is a significant positive correlation between perceived Visual Spatial intelligence and academic achievement of the students. It can also be concluded that this correlation is moderate. |
| 5 | Musical | 0.407 | 0.000 | There is a significant positive correlation between perceived Music intelligence and academic achievement of the students. It can also be concluded that this correlation is moderate. |
| 6 | Naturalist | 0.391 | 0.000 | There is a significant positive correlation between perceived Naturalist intelligence and academic achievement of the students. It can also be concluded that this correlation is weak. |
| 7 | Intrapersonal | 0.370 | 0.002 | There is a significant positive correlation between perceived intrapersonal intelligence and academic achievement of the students. It can also be concluded that this correlation is weak. |
| 8 | Logic-Mathematic | 0.264 | 0.001 | There is a significant positive correlation between perceived logic-mathematic intelligence and academic achievement of the students. It can also be concluded that this correlation is weak. |

## V. SUMMARY AND FUTURE PLAN

In conclusion, lecturers should be creative when designing their teaching materials or activities to enable students to use their intelligence in the classroom. With the interactive and suitable teaching materials and activities in class, student attention can be retained and also, improve lecturer's passion and motivation to teach and prepare teaching materials. Designing online MI teaching activities at higher education is the best solution to facilitate lecturers to create their own teaching materials without having any IT knowledge especially in programming. Lecturer can reduce preparation time and indirectly attract the attention of students to learn and use the materials effectively. It is recommended that an online MI teaching tool should be developed in order to achieve an effective teaching and learning process.

ACKNOWLEDGMENT

Siti Nurul Mahfuzah Mohamad would like to thank her primary advisor, Prof. Madya Dr. Sazilah Salam and Dr. Norasiken Bakar for their advice, support, encouragement and guidance. She also likes to express gratitude to the diploma students from Malaysia Polytechnic, who participated in this paper, as well as to all the Polytechnic lecturers for their support and cooperation. Thanks to the Malaysia Ministry of Higher Education (MOHE) who sponsor her study and deep appreciations are also dedicated to anyone who directly or indirectly involved in this study. This research is supported by UTeM (Universiti Teknikal Malaysia Melaka) under project no MTUN/2012/UTeM-FTMK/10 M00018.

REFERENCES

[1] Aris, B., Abu, M., Ellington, H. & Dhamotharan, M. (2000). *Learning About Information Technology In Education Using Multimedia*. In D. Willis et al. (Eds.), Proceedings of Society for Information Technology & Teacher Education International Conference 2000 (pp. 762-767).

[2] Armstrong. (1994). *Multiple Intelligences in the Classroom.* Virginia: ASCD. A nuts-and-bolts guide to multiple intelligences covering subjects such as lesson planning, teaching strategies, classroom management, activity centers, thematic instruction, assessment, special education, cognitive skills, and cultural diversity.

[3] Arvaja, M., Hakkinen, P., & Kankaanranta, M. (2008). *Collaborative Learning and Computer-Supported Collaborative Learning Environments*. In J. Voogt, & G. Knezek, International Handbook of Information Technology in Primary and Secondary Education (pp. 267-279). Springer US.

[4] Baum, S., Vines, 1., & Slatin, B. (2005). *Multiple intelligences in the elementary classroom: A teacher's toolkit.* New York, NY: Teachers College Press.

[5] Bushro, A. & Halimah B. Z. (2007). *Preliminary Testing on Interactive Multimedia Courseware for Mathematics based on the theory of Multiple Intelligences.* Proceedings of the International Conference on Electrical Engineering and Informatics Institut Teknologi Bandung, Indonesia June 17-19, 2007.

[6] Bushro, A, 2008. *Kejuruteraan Perisian Kursus Multimedia Matematik Berasaskan Model Kecerdasan Pelbagai (MI-MathS).* Tesis PhD. Universiti Kebangsaan Malaysia (In Malay).

[7] Campbell, L., Campbell, B., & Dickinson, D. (1996). *Teaching & learning through multiple intelligences.* Needham Heights, MA: Allyn & Bacon.

[8] Che, M. Y. and M.N. Mariani, 2001. *Personaliti pelajar pintar cerdas dan hubungannya dengan pencapaian akademik*. Prosiding Konferensi Kebangsaan Kajian Pasca Siswazah. Serdang: Universiti Putra Malaysia (In Malay).

[9] Gardner, H. (1983). Frames of mind : *The theory of multiple intelligences.* Basic Books, INC., Publishers, New York.

[10] Gardner, H. (1993). *Frames of mind: the theory of multiple intelligences* (2nd ed.). London: Fontana Press.

[11] Gardner, H. 2000. *The Disciplined Mind: Beyond Facts and Standardized Tests, the K-12 Education that Every Child Deserves*. NY: Penguin Books.

[12] Gardner, H. (2001). The Three Faces of Intelligence.

[13] Gardner, H. (2006). *Multiple intelligences: New* horizons. New York: Basic Books.

[14] Ghazi, S. R., G. Shahzada, U.S. Gilani, M.N. Shabbir and M. Rashid, 2011. *Relationship Between Student's Self Perceived Multiple Intelligences And Their Academic Achievement*. International Journal of Academic Research. Part II. 3(2): 619-623.

[15] Hajimirzayee, F. and M.K.S. Abadi, 2012. *The Relationship between Iranian EFL Students' Knowledge of Farsi Grammar and their English Grammar Knowledge.* Journal of Academic and Applied Studies. 2(10):16-33.

[16] Laidra, K., H. Pullmann and J. Allik, 2007. *Personality and intelligence as predictors of academic achievement:* A cross-sectional study from elementary to secondary school. Personality and Individual Differences, 42(3): 441-451.

[17] Lazear, D. (1994). *Seven pathways of learning: teaching students and parents about multiple intelligences*. Tucson, AZ: Zephyr Press.

[18] Mostafari, B.B.M., O, Akbari and F. Masoominezhad, 2012. *The Relationship between Interpersonal, Visual-Spatial Intelligences and Technical Translation Quality.* 2(8):176-184.

[19] McInnerney, J. M., & Roberts, T. S. (2004). *Collaborative or Cooperative Learning? In T. S.Roberts, Online Collaborative Learning, Theory And Practice* (pp. 203-214). Central Queensland University, Australia: Idea Group Publishing.

[20] Moran, S., M.Kornhaber and Gardner, H, 2006. *Orchestrating multiple intelligences.* Educational Leadership. National Commission on Excellence in Education. A Nation at Risk: The Imperative for Educational Reform. Washington, D.C. United States Department of Education.

[21] Ozdemir, P., Giineysu, S., & Tekkaya, C. (2006). *Enhancing learning through multiple intelligence*. [Electronic version]. Journal of Biological Education, 40(2), 74-78.

[22] Parviz, A. and S. Farhady, 2012. *Using Differentiated Instruction to Teach Vocabulary in Mixed Ability Classes with a Focus on Multiple Intelligences and Learning Styles*. International Journal of Applied Science and Technology. 2(4):72-82.

[23] Rettig, M, 2005. *Using the multiple intelligences to enhance instruction for young children and young children with disabilities.* Early Childhood Education Journal, 32: 255-259.

[24] Smith, B. L., & MacGregor, J. T. (1992). *What is Collaborative Learning?* In A. S. Goodsell,M. R. Maher, & V. Tinto, Collaborative Learning - A Sourcebook For Higher Education (pp.10-30). National Center on Post Secondary Teaching, Learning And Assessment, Syracuse University

[25] Suzanna Ganggi. (2011). *Differentiating Instruction using Multiple Intelligences in the Elementary School Classroom* : A Literature Review. Master Thesis. University of Wisconsin-Stout, Menomonie, United States.

[26] Toth M.J., Audrey A. B, Foulger T. S. (2010). *Changing Delivery Methods, Changing Practices: Exploring Instructional Practices in Face-to-Face and Hybrid Courses.* MERLOT Journal of Online Learning and Teaching Vol. 6, No. 3.

[27] Tyler, C.E., (2011). *Can Multiple Intelligences Enhance Learning for Higher Education On-Line Instruction?.* E-Leader Vietnam.

[28] Waterhouse, L, 2006. *Multiple intelligences, the Mozart effect, and emotional intelligence: A critical review*. Educational Psychologist, 41: 207–225.

[29] Wilson, L. (1998). *Multiple intelligences defined.* Retrieved from http://www.uwsp.edu/education/lwilson/

[30] Woolfolk, A. (1990). *Educational psychology* (4th ed.). Englewood Cliffs, NJ: Prentice-Hall, (622 pp.).